\RequirePackage[pagewise,mathlines]{lineno}
\documentclass[twocolumn,aps,showpacs,superscriptaddress]{revtex4}
\usepackage{graphicx}
\usepackage{dcolumn}
\usepackage{bm}
\begin{document}

\def\Journal#1#2#3#4{{#1} {\bf #2}, #3 (#4)}

\def\NCA{Nuovo Cimento}
\def\NIM{Nucl. Instr. Meth.}
\def\NIMA{{Nucl. Instr. Meth.} A}
\def\NPB{{Nucl. Phys.} B}
\def\NPA{{Nucl. Phys.} A}
\def\PLB{{Phys. Lett.}  B}
\def\PRL{Phys. Rev. Lett.}
\def\PRC{{Phys. Rev.} C}
\def\PRD{{Phys. Rev.} D}
\def\ZPC{{Z. Phys.} C}
\def\JPG{{J. Phys.} G}
\def\CPC{Comput. Phys. Commun.}
\def\EPJ{{Eur. Phys. J.} C}
\def\PR{Phys. Rept.}
\def\PRV{Phys. Rev.}
\def\JHEP{JHEP}

\preprint{}
\title{Di-electron spectrum at mid-rapidity in $p+p$ collisions at $\sqrt{s} = 200$ GeV}
\affiliation{AGH University of Science and Technology, Cracow,
Poland} \affiliation{Argonne National Laboratory, Argonne,
Illinois 60439, USA} \affiliation{Brookhaven National Laboratory,
Upton, New York 11973, USA} \affiliation{University of California,
Berkeley, California 94720, USA} \affiliation{University of
California, Davis, California 95616, USA} \affiliation{University
of California, Los Angeles, California 90095, USA}
\affiliation{Universidade Estadual de Campinas, Sao Paulo, Brazil}
\affiliation{Central China Normal University (HZNU), Wuhan 430079,
China} \affiliation{University of Illinois at Chicago, Chicago,
Illinois 60607, USA} \affiliation{Krakow Universitiy of
Technology, Crakow, Poland} \affiliation{Creighton University,
Omaha, Nebraska 68178, USA} \affiliation{Czech Technical
University in Prague, FNSPE, Prague, 115 19, Czech Republic}
\affiliation{Nuclear Physics Institute AS CR, 250 68
\v{R}e\v{z}/Prague, Czech Republic} \affiliation{University of
Frankfurt, Frankfurt, Germany} \affiliation{Institute of Physics,
Bhubaneswar 751005, India} \affiliation{Indian Institute of
Technology, Mumbai, India} \affiliation{Indiana University,
Bloomington, Indiana 47408, USA} \affiliation{Alikhanov Institute
for Theoretical and Experimental Physics, Moscow, Russia}
\affiliation{University of Jammu, Jammu 180001, India}
\affiliation{Joint Institute for Nuclear Research, Dubna, 141 980,
Russia} \affiliation{Kent State University, Kent, Ohio 44242, USA}
\affiliation{University of Kentucky, Lexington, Kentucky,
40506-0055, USA} \affiliation{Institute of Modern Physics,
Lanzhou, China} \affiliation{Lawrence Berkeley National
Laboratory, Berkeley, California 94720, USA}
\affiliation{Massachusetts Institute of Technology, Cambridge, MA
02139-4307, USA} \affiliation{Max-Planck-Institut f\"ur Physik,
Munich, Germany} \affiliation{Michigan State University, East
Lansing, Michigan 48824, USA} \affiliation{Moscow Engineering
Physics Institute, Moscow Russia} \affiliation{Ohio State
University, Columbus, Ohio 43210, USA} \affiliation{Old Dominion
University, Norfolk, VA, 23529, USA} \affiliation{Panjab
University, Chandigarh 160014, India} \affiliation{Institute of
Nuclear Physics PAS, Cracow, Poland} \affiliation{Pennsylvania
State University, University Park, Pennsylvania 16802, USA}
\affiliation{Institute of High Energy Physics, Protvino, Russia}
\affiliation{Purdue University, West Lafayette, Indiana 47907,
USA} \affiliation{Pusan National University, Pusan, Republic of
Korea} \affiliation{University of Rajasthan, Jaipur 302004, India}
\affiliation{Rice University, Houston, Texas 77251, USA}
\affiliation{Universidade de Sao Paulo, Sao Paulo, Brazil}
\affiliation{University of Science \& Technology of China, Hefei
230026, China} \affiliation{Shandong University, Jinan, Shandong
250100, China} \affiliation{Shanghai Institute of Applied Physics,
Shanghai 201800, China} \affiliation{SUBATECH, Nantes, France}
\affiliation{Texas A\&M University, College Station, Texas 77843,
USA} \affiliation{University of Texas, Austin, Texas 78712, USA}
\affiliation{University of Houston, Houston, TX, 77204, USA}
\affiliation{Tsinghua University, Beijing 100084, China}
\affiliation{United States Naval Academy, Annapolis, MD 21402,
USA} \affiliation{Valparaiso University, Valparaiso, Indiana
46383, USA} \affiliation{Variable Energy Cyclotron Centre, Kolkata
700064, India} \affiliation{Warsaw University of Technology,
Warsaw, Poland} \affiliation{University of Washington, Seattle,
Washington 98195, USA} \affiliation{Wayne State University,
Detroit, Michigan 48201, USA} \affiliation{Yale University, New
Haven, Connecticut 06520, USA} \affiliation{University of Zagreb,
Zagreb, HR-10002, Croatia}

\author{L.~Adamczyk}\affiliation{AGH University of Science and Technology, Cracow, Poland}
\author{G.~Agakishiev}\affiliation{Joint Institute for Nuclear Research, Dubna, 141 980, Russia}
\author{M.~M.~Aggarwal}\affiliation{Panjab University, Chandigarh 160014, India}
\author{Z.~Ahammed}\affiliation{Variable Energy Cyclotron Centre, Kolkata 700064, India}
\author{A.~V.~Alakhverdyants}\affiliation{Joint Institute for Nuclear Research, Dubna, 141 980, Russia}
\author{I.~Alekseev}\affiliation{Alikhanov Institute for Theoretical and Experimental Physics, Moscow, Russia}
\author{J.~Alford}\affiliation{Kent State University, Kent, Ohio 44242, USA}
\author{B.~D.~Anderson}\affiliation{Kent State University, Kent, Ohio 44242, USA}
\author{C.~D.~Anson}\affiliation{Ohio State University, Columbus, Ohio 43210, USA}
\author{D.~Arkhipkin}\affiliation{Brookhaven National Laboratory, Upton, New York 11973, USA}
% ...

\collaboration{STAR Collaboration}\noaffiliation

\date{\today}
\begin{abstract}
We report on mid-rapidity mass spectrum of di-electrons and cross
sections of pseudoscalar and vector mesons via $e^{+}e^{-}$
decays, from $\sqrt{s} = 200$ GeV $p+p$ collisions, measured by
the large acceptance experiment STAR at RHIC. The ratio of the
di-electron continuum to the combinatorial background is larger
than 10\% over the entire mass range. Simulations of di-electrons
from light-meson decays and heavy-flavor decays (charmonium and
open charm correlation) are found to describe the data. The
extracted $\omega\rightarrow e^{+}e^{-}$ invariant yields are
consistent with previous measurements. The mid-rapidity yields
($dN/dy$) of $\phi$ and $J/\psi$ are extracted through their
di-electron decay channels and are consistent with the previous
measurements of $\phi\rightarrow K^{+}K^{-}$ and
$J/\psi\rightarrow e^{+}e^{-}$. Our results suggest a new upper
limit of the branching ratio of the $\eta \rightarrow e^{+}e^{-}$
of $1.7\times10^{-5}$ at 90\% confidence level.
\end{abstract}
\pacs{25.75.Dw} \maketitle

\section{Introduction}
Di-leptons are a crucial probe of the strongly interacting matter
created in ultra-relativistic heavy-ion
collisions~\cite{starwhitepaper,otherwhitepapers}. Leptons are
produced during the whole evolution of the created matter and can
traverse the medium with minimal interactions. Different
kinematics of di-lepton pairs (mass and transverse momentum
ranges) can selectively probe the properties of the formed matter
throughout its entire evolution~\cite{dilepton,dileptonII}.

In the low invariant mass range of produced lepton pairs
($M_{ll}\!<\!1.1$ GeV/$c^{2}$), vector meson in-medium properties
(mass and width of the spectral functions of $\rho(770),
\omega(782)$, and $\phi(1020)$) may be studied via di-lepton
decays and may exhibit modifications related to possible chiral
symmetry restoration~\cite{dilepton,dileptonII}. For example, at
the SPS, an explanation of the low-mass di-lepton enhancement in
the CERES $e^+e^-$ data from Pb+Pb collisions requires substantial
medium effects on the $\rho$-meson spectral function~\cite{ceres}.
Also, NA60 recently reported a significant excess of low-mass
$\mu^+\mu^-$ pairs in In+In collisions above the yield expected
from neutral meson decays~\cite{na60} which is consistent with a
broadened spectral function~\cite{massbroaden} but not a
dropping-mass scenario~\cite{dropmass}.

At RHIC, the PHENIX experiment observed a significant enhancement
for $0.15\!<M_{ee}\!<\!0.75$ GeV/$c^{2}$ in the low transverse
momentum ($p_{T}\!<\!1$ GeV/$c$) part of the di-electron continuum
in Au+Au collisions compared to that expected from hadronic
sources~\cite{lowmass}. Models that successfully describe the SPS
di-lepton data consistently fail to describe the PHENIX data in
the low-mass and low-$p_T$ region~\cite{lowmass,rapp:09}. Also, in
the higher $p_T$ range, direct photon yields were derived through
di-electron measurements at RHIC allowing an assessment of thermal
radiation~\cite{thermalphoton}. Additional precision experiments
with large acceptance and a broad range of beam energies can
provide invaluable insights in this subject~\cite{starwhitepaper}.

The di-lepton spectra in the intermediate mass range
($1.1\!<M_{ll}\!<\!3.0$ GeV/$c^{2}$) are expected to be directly
related to the thermal radiation of the Quark-Gluon Plasma
(QGP)~\cite{dilepton,dileptonII}. However, significant background
contributions from other sources have to be measured
experimentally. Such contributions include background pairs from
correlated open heavy-flavor decays, which produce a pair of
electrons or muons from the semileptonic decay of a pair of open
charm or bottom hadrons ($c\bar{c}\rightarrow l^{+}l^{-}$ or
$b\bar{b}\rightarrow l^{+}l^{-}$). In the high-mass region
($M_{ll}\!>\!3.0$ GeV/$c^{2}$), $J/\psi, \Upsilon$, and their
excited states are used to study the color screening features of
the QGP~\cite{jpsitheory}. The PHENIX collaboration has reported
di-electron spectrum in p+p collisions and has found the data are
very well described by the hadronic cocktail and heavy flavor
decays for the entire mass range within the uncertainty of the
data and the cocktail~\cite{lowmass}. The first di-electron
continuum measurement from STAR in $\sqrt{s} = 200$ GeV $p+p$
collisions, presented in this paper, provides a crucial reference
for corresponding future STAR measurements in heavy-ion
collisions.

Rare processes like leptonic decays of hadrons provide possible
observables to be used in searching for traces of new physics
Beyond the Standard Model
(BSM)~\cite{dorokhov,bjorken,savage,Bergstrom}. These decays
usually involve electromagnetic or weak couplings which can be
calculated to a high degree of accuracy within the Standard Model
(SM). In addition to a direct observation of the Higgs boson, the
Large Hadron Collider (LHC) looks to explore BSM physics.
Deviations of rare process observables from SM predictions may be
taken as indirect evidence of a new coupling beyond the SM
physics~\cite{dorokhov}, which can also be explored at lower
energies. The pseudoscalar mesons (for example, $\eta$ or $\eta' $
) are particularly interesting since their decay to $e^+ e^-$
pairs is suppressed by $\alpha^2 (10^{-4})$ and by helicity
conservation due to the small electron mass ($r^2 = (m_e/m_\eta)^2
\simeq 10^{-6})$~\cite{etaprime}. The branching ratio (B.R.) of
$\eta\rightarrow e^{+}e^{-}$ is $2.3\times10^{-9}$ according to
the SM predictions, however, couplings from BSM physics may
increase this B.R. significantly~\cite{dorokhov}. RHIC offers
high-luminosity nucleus-nucleus collisions with large
multiplicities and copious hadrons of interest thereby providing a
unique environment for studying rare decay processes, nuclear
medium effects, and searching for BSM physics. With recent
upgrades, including the new Time-Of-Flight detector
(TOF)~\cite{startof} and improved data acquisition
system~\cite{stardaq}, the STAR experiment is able to benefit from
a high rate capability as well as excellent lepton identification
at low momentum in the search for rare decays.

This paper is organized as follows. Section~\ref{detector} shows
the detector and data sample used in this analysis.
Section~\ref{analysis} describes the analysis details including
electron identification, electron-pair distributions, background
subtraction, and di-electron continuum without efficiency
correction. Section~\ref{dieresults} presents the details of the
simulations of di-electrons from light-meson decays and
heavy-flavor decays, collectively called cocktail simulations. The
efficiency correction for the di-electron continuum, the corrected
di-electron continuum, and systematic uncertainties are also
discussed in this section. Results on the yields of $\omega$,
$\phi$, and $J/\psi$ from di-electronic decays are presented in
detail in Sec.~\ref{vecresults}. The rare decay of $\eta
\rightarrow e^{+}e^{-}$ is discussed in Sec.~\ref{raredecay}.
Lastly, Sec.~\ref{summary} provides a concluding summary.

\section{Detectors And Data Sample}\label{detector} Two sub-detectors
are used in this analysis at mid-rapidity at STAR~\cite{star}: the
Time Projection Chamber (TPC)~\cite{startpc} and a newly installed
TOF~\cite{startof}. The TPC is the main tracking detector at STAR,
measuring momenta, charge, and energy loss of charged particles.
The TPC, immersed in 0.5 Tesla solenoidal magnetic field, is a 4.2
m long cylinder surrounding the beam pipe. Its fiducial volume
ranges from 50 to 200 cm radially and is filled with P10 gas (90\%
argon and 10\% methane). Electrons from ionized gas drift toward
the nearest endcap, where the signals are read out. The TPC
readout is divided azimuthally into 24 sectors, 12 at each end.
Each sector is divided into inner and outer subsectors with a
total of 45 pad row readouts. The pad row readouts provide precise
positions of primary ionization clusters (hits) generated by
charged particles. The ionization energy loss of a charged
particle in the TPC gas ($dE/dx$) is used for particle
identification~\cite{bichsel,pidpp08}. The most probable $dE/dx$
is determined from the mean of a distribution in which the 30\% of
clusters with the largest signals are discarded (i.e. a truncated
mean). For the data taken in 2009 and analyzed here, 72\% of the
full TOF system was installed and operational. The full TOF system
contains 120 units which we call trays, 60 in the pseudo-rapidity
range $0\!<\!\eta\!<\!0.9$ and 60 for $-0.9\!<\!\eta\!<\!0$, with
each tray covering 6 degrees in azimuth. The TOF has a typical
stop timing resolution of 85 ps allowing the identification of
$\pi$($K$) up to a momentum of 1.6 GeV/$c$ and $p(\bar{p})$ up to
a momentum of 3 GeV/$c$~\cite{pidNIMA,tofPID}.

The minimum-bias triggered events were defined by the coincidence
of signals in the two Vertex Position Detectors
(VPDs)~\cite{starvpd}, located on each side of the STAR
interaction region and covering $4.4<|\eta|<4.9$. This di-electron
analysis used 107 million minimally biased events from non-singly
diffractive (NSD) $\sqrt{s} = 200$ GeV $p+p$ collisions
($\sigma_{_{NSD}}=30.0\pm3.5$ mb~\cite{starhighpt}), in which the
collision vertex is required to be within 50 cm of the mean of the
distribution, nominally at the center of the TPC and along the
beam line.

\section{Data Analysis}\label{analysis}
\subsection{Tracking and particle identification}
Hits belonging to charged particles traversing the TPC are
collected and reconstructed into tracks with well defined
geometry, momentum ($p$), and {\it dE/dx}. Only tracks that
project back to within 1 cm of the collision vertex are retained
in the analysis, thereby limiting the combinatorial background
from conversions and enabling a high detecting efficiency. The
tracks are required to have at least 25 hits out of a maximum of
45 to avoid split tracks. Also, a minimum of 15 hits is required
in the {\it dE/dx} measurement to obtain good $dE/dx$
resolution~\cite{startpc,pidlowpt}. For particles directly
originating from the collision, the collision vertex is added as
an additional hit to further improve the momentum
measurement~\cite{pidlowpt}.

\begin{figure*}[htbp]
\begin{center}
\includegraphics[keepaspectratio,width=0.9\textwidth]{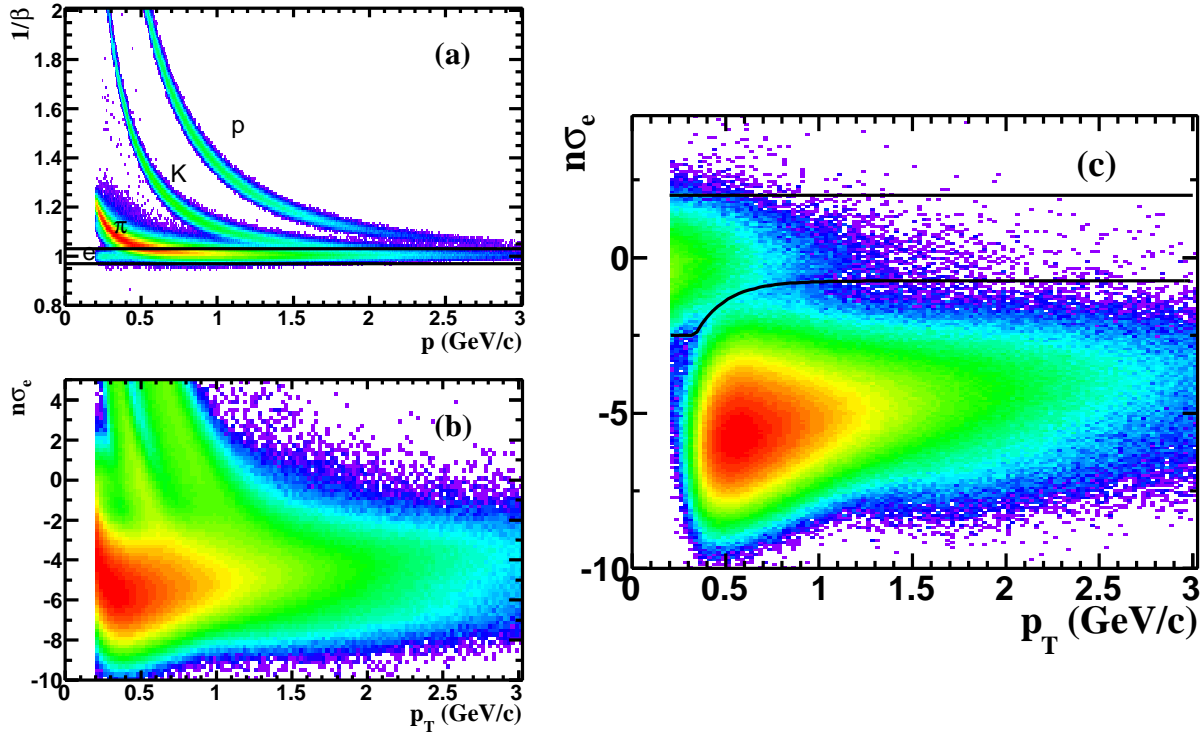}
\caption{(Color online) Panel (a): $1/\beta$ versus momentum of
tracks from the TOF with $|\eta|\!<\!1$ from 200 GeV $p+p$
collisions. The line indicates the cut of $|1/\beta-1|\!<\!0.03$.
Panel (b): The normalized $dE/dx$ distribution from the TPC as a
function of $p_T$. Panel (c): The normalized $dE/dx$ distribution
from the TPC as a function of $p_T$ with the cut of
$|1/\beta-1|\!<\!0.03$. An electron band is prominent, indicated
by the lines, with the requirement of velocity close to the speed
of light from TOF measurement. } \label{pid}
\end{center}
\end{figure*}

\begin{figure}[htbp]
\begin{center}
\includegraphics[keepaspectratio,width=0.45\textwidth]{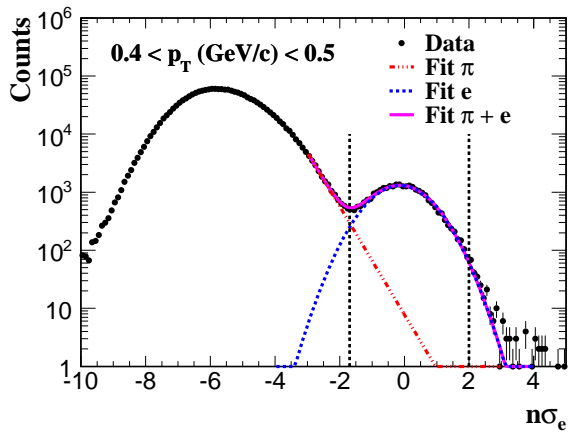}
\caption{(Color online) The $n\sigma_{e}$ distribution in
$0.4\!<\!p_T\!<\!0.5$ GeV/$c$ after the cut of
$|1/\beta-1|\!<\!0.03$ is applied. The solid curve represents a
Gaussian plus exponential fit to the $n\sigma_{e}$ distribution.
The dot-dashed line is for the hadron component and the dashed is
for the electron contribution. The two dashed lines perpendicular
to the x-axis represent the range of the $n\sigma_{e}$ cut to
ensure a high purity for electron candidates in this $p_T$ range.
} \label{eprojection}
\end{center}
\end{figure}

\begin{figure}[htbp]
\begin{center}
\includegraphics[keepaspectratio,width=0.45\textwidth]{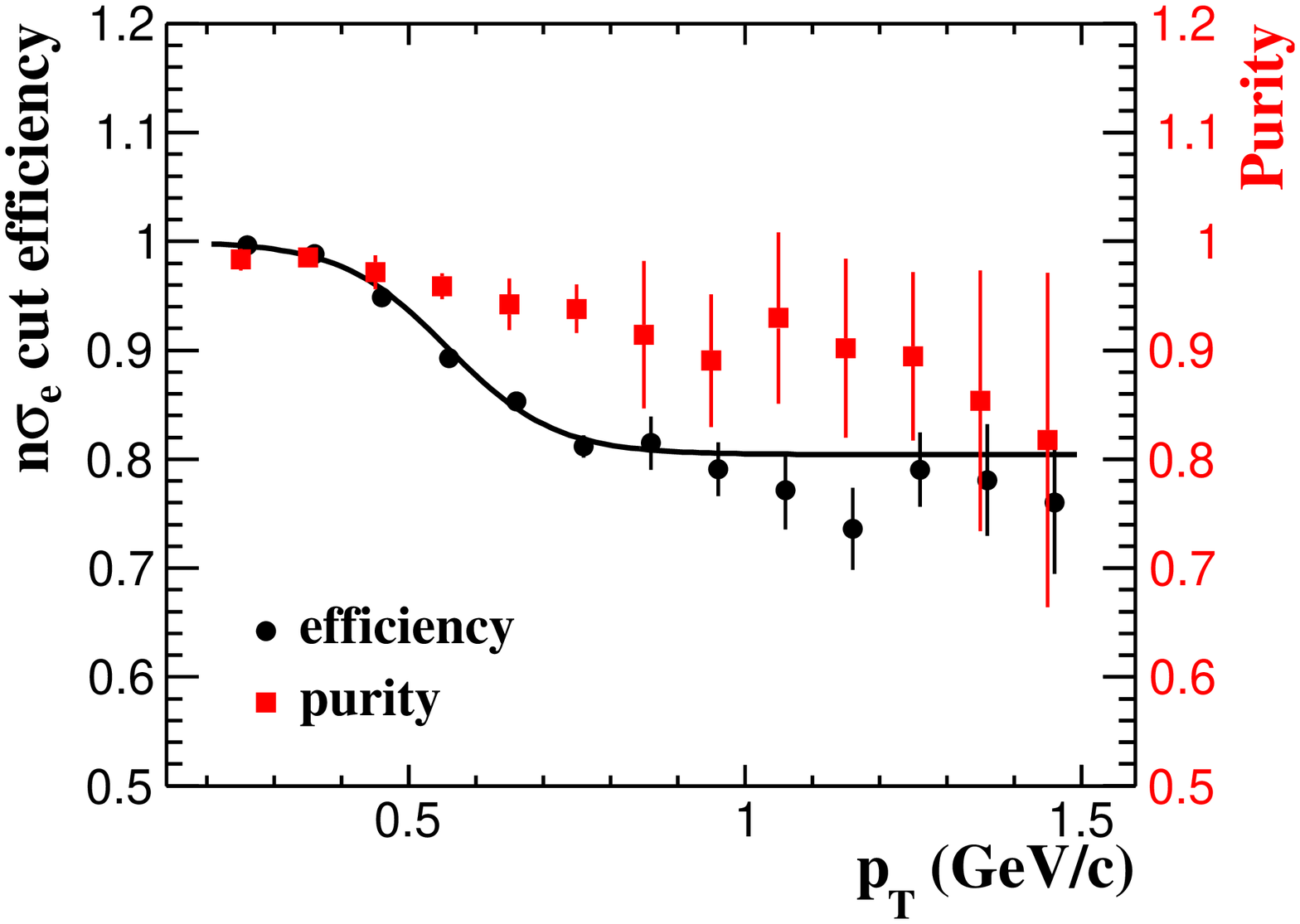}
\caption{(Color online) The purity and the $n\sigma_{e}$ cut
efficiency for electron candidates as a function of $p_T$ in
$|\eta|\!<\!1$ for $p+p$ collisions at $\sqrt{s} = 200$ GeV. The
squares represent the purity and closed circles represent the
$n\sigma_{e}$ efficiency. The $p_T$ positions of the last three
data points for the efficiency are slightly shifted for clarity.
The error bars are the quadrature sum of statistical and
systematic uncertainties. A function form of $A+B/[C+exp(D\times
p_{T})]$ is used to fit the efficiency data points and for the
efficiency correction, in which $A$, $B$, $C$ and $D$ are the fit
parameters. A constant component in the fit for $p_T\!>\!0.8$
GeV/$c$ is driven by the study in Ref.~\cite{pidpp08}. }
\label{dedxeff}
\end{center}
\end{figure}

Figure~\ref{pid} panel (a) shows the $1/\beta$ from TOF of
particles versus momentum in $p+p$ collisions while panels (b) and
(c) show the normalized $dE/dx$ ($n\sigma_{e}$) distribution from
the TPC as a function of $p_T$, without and with a requirement of
high velocity $|1/\beta-1|\!<\!0.03$, respectively. The quantity
$n\sigma_{e}$ is defined as: $n\sigma_{e}=\ln(dE/dx/I_{e})/R_{e}$,
where $dE/dx$ is the measured specific energy loss of a particle,
and $I_{e}$ is the expected $dE/dx$ of an electron. $R_{e}$ is the
$\ln(dE/dx/I_{e})$ resolution of an electron and is better than
8\%. Electron candidates whose $n\sigma_{e}$ falls between the
lines indicated in Figure~\ref{pid} panel (c) are retained in this
analysis. With a perfect calibration, the $n\sigma_{e}$ for single
electrons should follow a standard normal distribution.
\renewcommand{\floatpagefraction}{0.75}
Figure~\ref{eprojection} shows the $n\sigma_{e}$ distribution for
$0.4\!<\!p_T\!<\!0.5$ GeV/$c$ after the cut of
$|1/\beta-1|\!<\!0.03$. The two dashed lines perpendicular to the
x-axis represent the range of the $n\sigma_{e}$ cut in this $p_T$
region. A Gaussian plus exponential function, respectively
representing the electron and hadron components, is used to fit
the $n\sigma_{e}$ distribution. From the fit, we derive the purity
and the $n\sigma_{e}$ cut efficiency on electron candidates as a
function of $p_T$, as shown in Fig.~\ref{dedxeff}. The purity is
defined within a range of $n\sigma_{e}$ (i.e. between the vertical
dashed lines in Fig.~\ref{eprojection}) as being the ratio of the
electron counts in the area of the dashed Gaussian to the counts
of all particles. The efficiency is defined to be the ratio of the
electron counts under the dashed Gaussian within a range of
$n\sigma_{e}$ to the total electron counts under the dashed
Gaussian. The errors on the efficiency and purity are determined
by adjusting the fit range. The electron yields are sensitive to
the fit range since the hadron contamination increases in the
smaller $n\sigma_{e}$ region thereby leading to large errors for
the efficiency for $p_T\!>\!0.8$ GeV/$c$. Our exponential
extrapolation of the $n\sigma_{e}$ distribution for the hadron
component tends to over-estimate the background and therefore
should be taken as an upper limit on the hadron contamination. By
combining the velocity ($\beta$) information from the TOF and the
$dE/dx$ from the TPC, electrons can be clearly identified from low
to intermediate $p_T$ ($p_T\!<\!3$ GeV/$c$) for
$|\eta|\!<\!1$~\cite{starelectron}.

\renewcommand{\floatpagefraction}{0.75}

\renewcommand{\floatpagefraction}{0.75}

\subsection{Di-electron invariant mass distribution and background subtraction}\label{bg}
With a high purity for the electron samples, the $e^{+}$ and
$e^{-}$ from the same events are combined to generate the
invariant mass distributions ($M_{ee}$) of $e^{+}e^{-}$ pairs
called unlike-sign distributions. The unlike-sign distributions
contain both signal and background. The signals are di-electrons
from light-meson decays and heavy-flavor decays (charmonium and
open charm correlation). The background results from random
combinatorial pairs and correlated pairs. Electron candidates are
required to be in the range of $|\eta|\!<1$ and $p_T>0.2$ GeV/$c$
while $e^{+}e^{-}$ pairs are required to be in the rapidity range
of $|y_{ee}|\!<\!1$.

The following two techniques are used for background estimation.
In the like-sign technique, electron pairs with the same charge
sign are combined from the same events. In the mixed-event
technique, unlike-sign pairs are formed from different events. In
order to ensure the events used in mixing have similar geometric
acceptance in the detector, we only mix events which have
collision vertices within 5 cm of each other along the beam line
direction.

Neither method represents the background perfectly. In the
low-mass region, there is a correlated cross pair background
(coming from two pairs of $e^{+}e^{-}$ from the same meson decays:
Dalitz decays followed by a conversion of the decay photon, or
conversions of multiple photons from the same meson). This
background is present in the like-sign distribution but not in the
mixed-event background. On the other hand, due to the sector
structure of detectors and the different curvatures of positively
and negatively charged particle tracks in the plane perpendicular
to the magnetic field, like-sign and unlike-sign pairs will have
different acceptance. Moreover, in the high invariant mass range,
there may be contributions from jet correlations which are absent
from the mixed-event technique.

\renewcommand{\floatpagefraction}{0.75}
\begin{figure}[htbp]
\begin{center}
\includegraphics[keepaspectratio,width=0.45\textwidth]{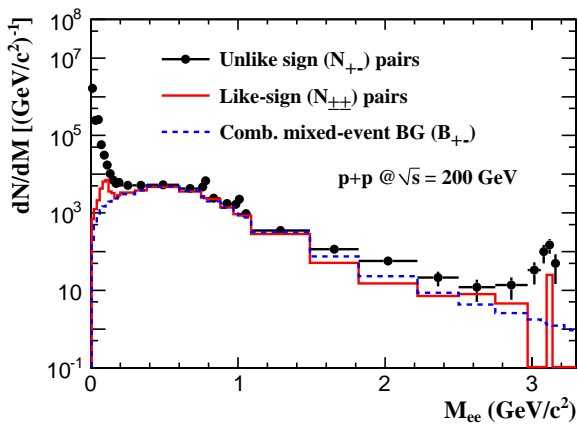}
\caption{(Color online) The electron-pair invariant mass
distributions for unlike-sign pairs, like-sign, and mixed-event
(combinatorial) background in minimum-bias $p+p$ collisions. The
electron candidates are required to be in the range of
$|\eta|\!<1$ and have a $p_T$ greater than 0.2 GeV/$c$. The $ee$
pairs were required to be in the rapidity range of
$|y_{ee}|\!<\!1$. The uneven bin widths are used based on the
yields and the signal to background ratios.} \label{eepair}
\end{center}
\end{figure}
Figure~\ref{eepair} shows the invariant mass distribution for
unlike-sign pairs, like-sign pairs, and mixed-event background.
The mixed-event distribution is normalized by a constant to match
the like-sign distribution in the mass range 0.4-1.5 GeV/$c^{2}$.
In our analysis we subtract the like-sign background for
$M_{ee}\!<\!0.4$ GeV/$c^{2}$ and the mixed-event background in the
higher-mass region.
\renewcommand{\floatpagefraction}{0.75}
\begin{figure*}[htbp]
\begin{center}
\includegraphics[keepaspectratio,width=0.9\textwidth]{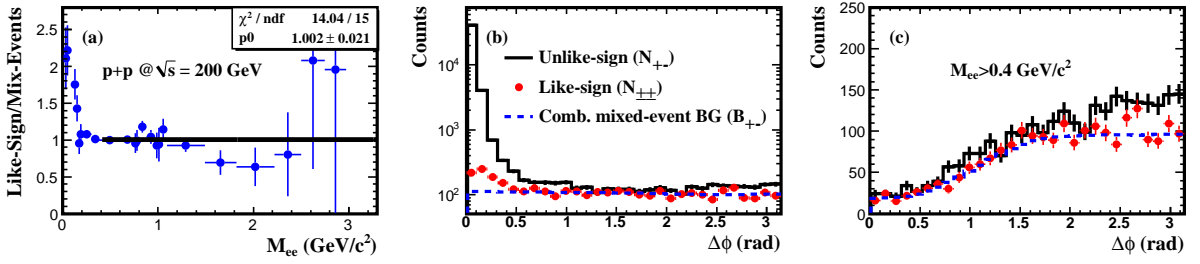}
\caption{(Color online) Panel (a): The ratio of like-sign to
mixed-event distributions in minimum-bias $p+p$ collisions. Panel
(b): The distribution of the difference of the azimuthal angles
($\Delta\phi$) of the two electrons in the unlike-sign, like-sign
and mixed-event pairs in minimum-bias $p+p$ collisions. Panel (c):
The $\Delta\phi$ distributions of unlike-sign, like-sign and
mixed-event pairs for $M_{ee}\!>\!0.4$ GeV/$c^{2}$ in minimum-bias
$p+p$ collisions. Errors are statistical.} \label{bgrat}
\end{center}
\end{figure*}
For a cross-check on the consistency of the two methods we compare
their shapes in the higher-mass region (Fig.~\ref{bgrat}). A
constant fits the ratio of like-sign over mixed-event
distributions for $M_{ee}\!>\!0.4$ GeV/$c^{2}$ with $\chi^{2}/NDF=
14/15$, as shown in the ratio plot in Fig.~\ref{bgrat} (a). We
also find that for electron pairs with significantly higher
transverse momentum (as determined from the Barrel
Electro-Magnetic Calorimeter triggered events), the shapes of
like-sign and mixed-event distributions agree in the mass region
of 1-3 GeV/$c^{2}$~\cite{ruan:11}.
Fig.~\ref{bgrat} (b) shows the distribution of the difference of
the azimuthal angles ($\Delta\phi$) of the two electrons in the
unlike-sign, like-sign and mixed-event pairs. The difference
between like-sign and mixed-event pairs at low $\Delta\phi$ can be
attributed to cross-pair contributions. For $M_{ee}\!>\!0.4$
GeV/$c^{2}$, the $\Delta\phi$ distributions of like-sign and
mixed-event pairs match nicely with each other as shown in
Fig.~\ref{bgrat} (c), indicating that mixed-event background
subtraction is valid in the corresponding mass region.

As an additional check with a different method, we also perform
the analysis by subtracting the like-sign background in the whole
mass region. The difference in the di-electron yields from the
mixed-event and like-sign methods is found to be within errors for
$M_{ee}\!>\!0.4$ GeV/$c^{2}$. We also correct for acceptance
differences between the like-sign and unlike-sign electron pairs
in both methods and will discuss the details in
Sec.~\ref{correctedcontinuum}.

\renewcommand{\floatpagefraction}{0.75}
\begin{figure*}[htbp]
\begin{center}
\includegraphics[keepaspectratio,width=0.9\textwidth]{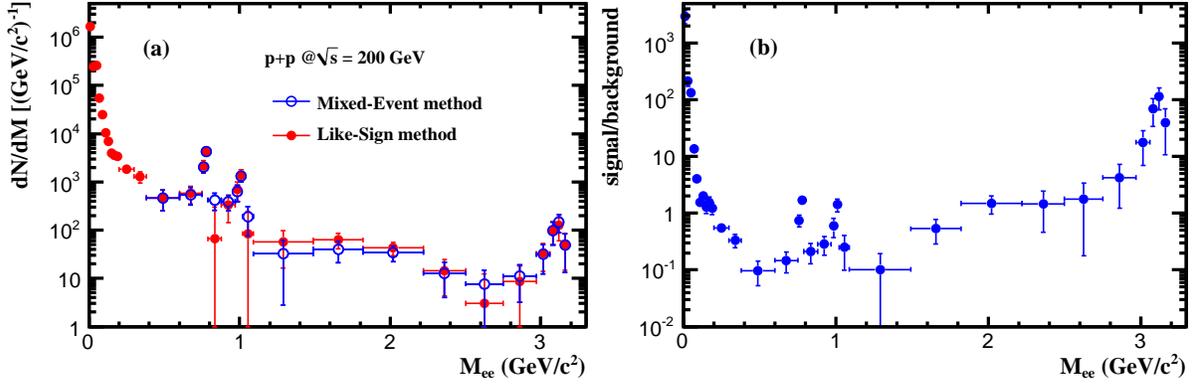}
\caption{(Color online) Panel (a): The di-electron ($e^{+}e^{-}$)
continuum after background subtraction without efficiency
correction in $\sqrt{s} = 200$ GeV minimum-bias $p+p$ collisions.
Two methods of obtaining the background are indicated. Errors are
statistical only. Panel (b): The signal over background ratio,
plotted as a function of $M_{ee}$ for NSD $p+p$ collisions at
$\sqrt{s} = 200$ GeV. Errors are statistical.}
\label{rawcontinuum}
\end{center}
\end{figure*}
Figure~\ref{rawcontinuum} (a) shows the di-electron continuum
after background subtraction from both like-sign and mixed-event
methods in $p+p$ collisions at $\sqrt{s} = 200$ GeV. The
measurements are done requiring $|y_{e^{+}e^{-}}|\!<1,
|\eta_{e}|\!<1$ and $p_T(e)\!>0.2$ GeV/$c$ and no efficiency
correction has been applied. The two methods give consistent
results for $M_{ee}\!>\!0.4$ GeV/$c^{2}$. For the following, we
use the yields from like-sign method for $M_{ee}\!<\!0.4$
GeV/$c^{2}$ and results obtained from the mixed-event method at
higher mass as the default since the mixed-event background
distribution matches the like-sign distribution and has better
precision for $M_{ee}\!>\!0.4$ GeV/$c^{2}$. The signal to
background ratio in $p+p$ collisions versus $M_{ee}$ is shown in
Fig.~\ref{rawcontinuum} (b).


\section{Di-electron continuum in STAR acceptance}\label{dieresults}


\subsection{Cocktail simulation}\label{cocktail} 
\begin{figure*}[htbp]
\begin{center}
\includegraphics[keepaspectratio,width=0.90\textwidth]{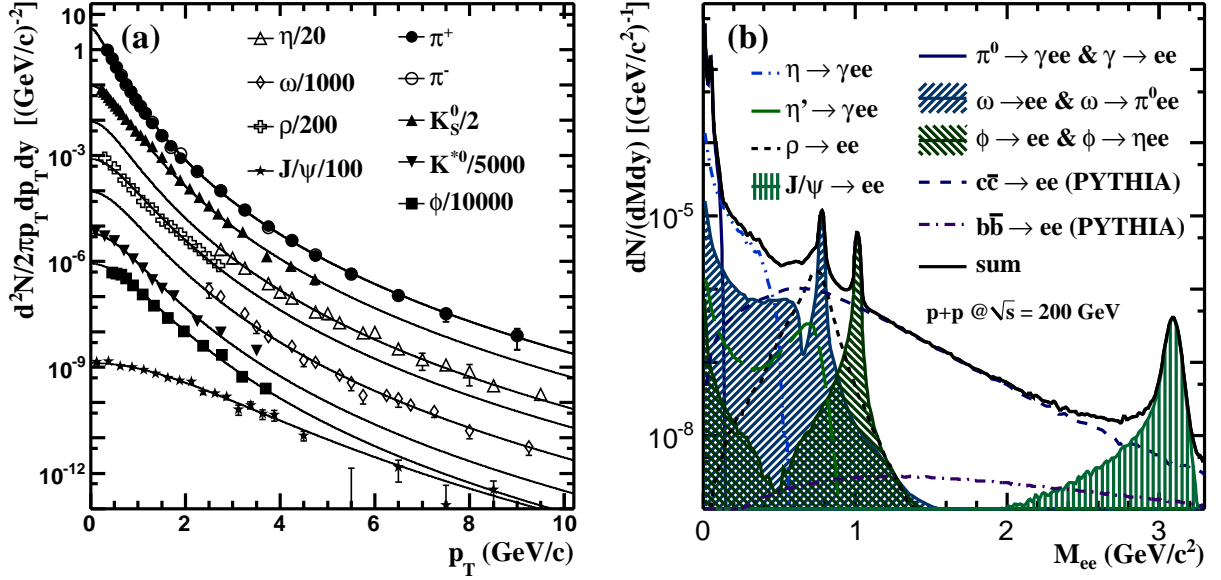}
\caption{(Color online) Panel (a): The invariant yields of
measured mesons fit with the Tsallis functions~\cite{Tsallis} in
$p+p$ collisions at $\sqrt{s} = 200$ GeV. See text for details.
Panel (b): The simulated raw di-electron continuum within STAR
acceptance for $\sqrt{s} = 200$ GeV minimum-bias $p+p$ collisions.
Different cocktail contributions are shown.} \label{mesonspectra}
\end{center}
\end{figure*}

The di-electron
pairs may come from decays of the light-flavor and heavy-flavor
hadrons. They include $\pi^{0}$, $\eta$, and $\eta^{\prime}$
Dalitz decays: $\pi^{0}\rightarrow \gamma e^{+}e^{-}$, $\eta
\rightarrow \gamma e^{+}e^{-}$, and $\eta^{\prime}\rightarrow
\gamma e^{+}e^{-}$; vector meson decays: $\omega \rightarrow
\pi^{0} e^{+}e^{-}$, $\omega \rightarrow e^{+}e^{-}$, $\rho^{0}
\rightarrow e^{+}e^{-}$, $\phi \rightarrow \eta e^{+}e^{-}$, $\phi
\rightarrow e^{+}e^{-}$, and $J/\psi \rightarrow e^{+}e^{-}$;
heavy-flavor hadron semi-leptonic decays: $c\bar{c} \rightarrow
e^{+}e^{-}$ and $b\bar{b} \rightarrow e^{+}e^{-}$; and Drell-Yan
contributions. We fit the invariant yields of mesons, previously
measured at STAR and PHENIX as discussed below, with the Tsallis
functions~\cite{Tsallis}, as shown in Fig.~\ref{mesonspectra} (a).
We use the Tsallis functions as input to a detector simulation in
which the particles are decayed into di-electrons with the
appropriate B.R.s. This GEANT detector simulation~\cite{geant}
uses a detailed description of the STAR geometry in 2009.
\renewcommand{\floatpagefraction}{0.75}
Simulated $e^{+}e^{-}$ cocktails from the various contributing
sources, as shown in Fig.~\ref{mesonspectra} (b), are selected
using the same cut conditions as those in the analyses of real
events. The di-electron contributions from the gamma conversion
$\gamma\rightarrow e^{+}e^{-}$ in the detector material are
accepted in both data and simulation subject to the same analysis
cuts as well. The imperfect description of the material in the
simulation leads to 3\% systematic uncertainty for the cocktail
simulation for $M_{ee}\!<\!0.1$ GeV/$c^2$. The Dalitz decays of
$\eta \rightarrow \gamma^{0} e^{+}e^{-}$, $\omega \rightarrow
\pi^{0} e^{+}e^{-}$ and $\eta^{\prime}\rightarrow \gamma
e^{+}e^{-}$ are obtained using the Kroll-Wada
expression~\cite{krollwada}.

For the Dalitz decays of $\pi^{0}$, $\eta$ and $\eta^{\prime}$, we
use the formula
\begin{equation}
{\frac{dN}{dM_{ee}}}\propto\sqrt{1-{\frac{4m_{e}^{2}}{M_{ee}^{2}}}}\bigg{(}1+{\frac{2m_{e}^{2}}{M_{ee}^{2}}}\bigg{)}{\frac{1}{M_{ee}}}\bigg{(}1-{\frac{M_{ee}^{2}}{M_{h}^{2}}}\bigg{)}^{3}|F(M_{ee}^{2})|^{2},
\end{equation} in which, $m_{e}$ is the electron mass, $M_{ee}$
is the di-electron mass, and $M_{h}$ is the mass of the hadron
which decays into the di-electron. $F(M_{ee}^{2})$ is the
electro-magnetic transition form factor. For the Dalitz decays of
vector mesons $\omega$ and $\phi$ ($A\rightarrow Be^{+}e^{-}$),
the formula is
\begin{widetext}
\begin{equation}
{\frac{dN}{dM_{ee}}}\propto\sqrt{1-{\frac{4m_{e}^{2}}{M_{ee}^{2}}}}\bigg{(}1+{\frac{2m_{e}^{2}}{M_{ee}^{2}}}\bigg{)}{\frac{1}{M_{ee}}}\bigg{[}\bigg{(}1+{\frac{M_{ee}^{2}}{M_{A}^{2}-M_{B}^{2}}}\bigg{)}^{2}
-{\frac{4M_{A}^{2}M_{ee}^{2}}{(M_{A}^{2}-M_{B}^{2})^{2}}}\bigg{]}^{3/2}|F(M_{ee}^{2})|^{2},
\end{equation}
\end{widetext} in which $M_{A}$ and $M_{B}$ are the mass of
particle A and B, respectively. For all Dalitz decays except
$\eta^{\prime}$, the form factor is parameterized as
\begin{equation}
F(M_{ee}^{2})={\frac{1}{1-M_{ee}^{2}\Lambda^{-2}}} ,
\end{equation} in which $\Lambda^{-2}$ is the form factor
slope. For $\eta^{\prime}$, we use the parametrization
from~\cite{etaprime}: \begin{equation}
|F(M_{ee}^{2})|^{2}={\frac{1}{(1-M_{ee}^{2}\Lambda^{-2})^{2}+\Gamma_{0}^{2}\Lambda^{-2}}}
,
\end{equation} where the $\Gamma_{0}^{2}$ is $1.99\times10^{-2}$ (GeV/$c^2)^{2}$.
The $\Lambda$ parameters are listed in Table~\ref{tab:I}.

The $\rho^{0} \rightarrow e^{+}e^{-}$ line shape is convoluted
with the Boltzmann phase space
factor~\cite{rholineshape,starrho0pp} and given by:

\begin{equation}
{\frac{dN}{dM_{ee}dp_{T}}}\propto{\frac{M_{ee}M_{\rho}\Gamma_{ee}}{(M_{\rho}^{2}-M_{ee}^{2})^{2}+M_{\rho}^{2}(\Gamma_{\pi\pi}+\Gamma_{ee}\Gamma_{2})^{2}}}\times
PS,
\end{equation}

\begin{equation}\Gamma_{\pi\pi}=\Gamma_{0}{\frac{M_{\rho}}{M_{ee}}}\bigg{(}{\frac{M_{ee}^{2}-4M_{\pi}^{2}}{M_{\rho}^{2}-4M_{\pi}^{2}}}\bigg{)}^{3/2},
\end{equation}

\begin{equation}\Gamma_{ee}=\Gamma_{0}{\frac{M_{\rho}}{M_{ee}}}\bigg{(}{\frac{M_{ee}^{2}-4m_{e}^{2}}{M_{\rho}^{2}-4m_{e}^{2}}}\bigg{)}^{1/2},
\end{equation}

\begin{equation}
PS={\frac{M_{ee}}{\sqrt{M_{ee}^{2}+p_{T}^{2}}}}exp\bigg{(}{-{\frac{\sqrt{M_{ee}^{2}+p_{T}^{2}}}{T}}}\bigg{)},
\end{equation}
where $M_{\rho}$ is 776 MeV/$c^2$, $M_{\pi}$ is the $\pi$ mass,
$\Gamma_{0}$ is 149 MeV/$c^2$, $\Gamma_{2}$ is the B.R. of
$\rho^{0} \rightarrow e^{+}e^{-}$, PS is the Boltzmann phase space
factor, and the inverse slope parameter $T$ is 160
MeV~\cite{rholineshape}. We neglect any contribution from the
interference effect~\cite{rhoomega} between $\rho^{0}$ and
$\omega$.

\begin{table*}\caption{The total yields at mid-rapidity ($dN/dy$) from the Tsallis fit,
decay B.R.s, and $\Lambda$ parameters of hadrons in NSD $p+p$
collisions at $\sqrt{s} = 200$ GeV.\label{tab:I}}
\begin{tabular}{c|c|c|c|c|c} \hline\hline
\centering
 meson & $\frac{dN}{dy}$ & relative uncertainty & decay channel & B.R. &$\Lambda^{-2}$ (GeV/$c^{2})^{-2}$\\
\hline
$\pi^{0}$   & $1.28$ & 14\% & $\gamma e^{+}e^{-}$ & $1.174\times10^{-2}$ & $1.756\pm0.022$~\cite{pdg}\\
$\eta$   & $1.7\times10^{-1}$ & 23\% & $\gamma e^{+}e^{-}$ & $7.0\times10^{-3}$&$1.95\pm0.18$~\cite{na60:09} \\
$\rho$   & $2.2\times10^{-1}$ & 15\% & $e^{+}e^{-}$ & $4.72\times10^{-5}$ &--\\
$\omega$   & $1.3\times10^{-1}$ & 21\% & $e^{+}e^{-}$ & $7.28\times10^{-5}$&-- \\
$\omega$   & & & $\pi^{0}e^{+}e^{-}$ & $7.7\times10^{-4}$ &$2.24\pm0.06$~\cite{na60:09}\\
$\phi$   & $1.7\times10^{-2}$ & 20\% & $e^{+}e^{-}$ & $2.954\times10^{-4}$ &--\\
$\phi$   & & & $\eta e^{+}e^{-}$ & $1.15\times10^{-4}$ &$3.8\pm1.8$~\cite{snd:01}\\
$\eta^{\prime}$   & $4.1\times10^{-2}$ & 29\%  & $\gamma e^{+}e^{-}$ & $4.7\times10^{-4}$~\cite{etaprimeBR}&$1.8\pm0.4$~\cite{etaprime} \\
$J/\psi$   & $2.4\times10^{-5}$ & 15\% & $e^{+}e^{-}$ & $5.94\times10^{-2}$ &--\\
\hline \hline
\end{tabular}
\end{table*}

The invariant yield of $\pi^{0}$ is taken as the average of
$\pi^{+}$ and $\pi^{-}$~\cite{tofPID,ppdAuPID}. The yields of
$\phi$~\cite{starphipp} and $\rho^{0}$~\cite{starrho0pp} are from
STAR while the $\eta$~\cite{phenixetapp},
$\omega$~\cite{phenixomegapp} and $J/\psi$~\cite{phenixjpsipp}
yields are measurements by PHENIX. Table~\ref{tab:I} lists the
total yields at mid-rapidity ($dN/dy|_{y=0}$) in 200 GeV NSD $p+p$
collisions.

The last input we consider in the cocktail simulation is the
$c\bar{c}$ cross section, which has been constrained by the
published low-$p_T$ $D^{0}$ spectrum in 200 GeV d+Au collisions
~\cite{starelectron}, the non-photonic electron spectrum in 200
GeV $p+p$ collisions~\cite{weihardprobes2010}, and the di-electron
continuum in this analysis. The details of these constraints will
be shown in Sec.~\ref{correctedcontinuum}. The $e^{+}e^{-}$ shapes
from open heavy-flavor pairs are obtained using PYTHIA6.416, in
which the $k_T$ factor is set by PARP(91)=1 GeV/c, and the parton
shower is set by PARP(67)=1~\cite{pythia}. With these parameters
chosen, PYTHIA can describe the shape of STAR measured $D^{0}$
spectrum and non-photonic electron spectrum. The total
contribution from the simulation is shown as the black solid curve
in Fig.~\ref{mesonspectra} (b). In the intermediate mass region,
the di-electron continuum is dominated by the $c\bar{c}$
contribution.


\subsection{Efficiency and acceptance correction}\label{eff} For
the di-electron continuum, the efficiency corrections are applied
within the STAR acceptance of $|y_{e^{+}e^{-}}|\!<\!1$,
$|\eta_{e}|\!<\!1$ and $p_T(e)\!>\!0.2$ GeV/$c$. The single
electron efficiency includes the TPC tracking efficiency, TOF
acceptance and detector response, and the $dE/dx$ efficiency.
Single electron tracking efficiency and acceptance in the TPC are
determined by Monte Carlo GEANT simulations. The TOF acceptance
and response efficiency for electrons is found to be 46\%
independent of $p_T$ for $|\eta|\!<\!1$~\cite{tofPID}. The
efficiency of the $n\sigma_{e}$ cut, used to ensure a high purity
for the electron sample, is close to 100\% at low $p_T$ and falls
to $\sim$80\% by $p_{T}=0.8$ GeV/c, as shown in
Fig.~\ref{dedxeff}.

\begin{figure}[htbp]
\begin{center}
\includegraphics[keepaspectratio,width=0.45\textwidth]{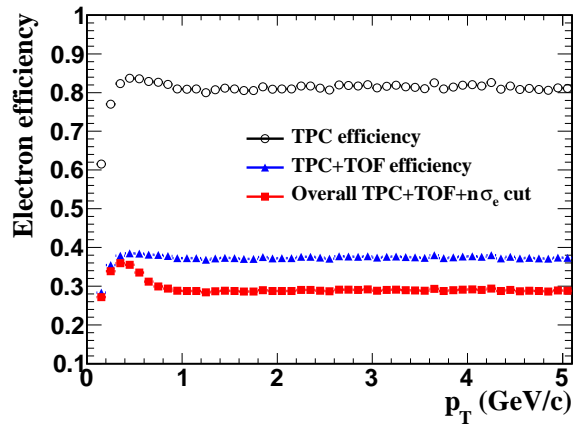}
\caption{(Color online) The efficiency for single electrons as a
function of $p_T$ in the pseudo-rapidity range of $|\eta|\!<\!1$
from $p+p$ collisions at $\sqrt{s} = 200$ GeV. Open circles
represent the TPC tracking efficiency. With additional TOF
matching and response, the efficiency is shown as triangles. With
additional $dE/dx$ cuts, the final efficiency is shown as squares.
} \label{eeff}
\end{center}
\end{figure}

\begin{figure}[htbp]
\begin{center}
\includegraphics[keepaspectratio,width=0.5\textwidth]{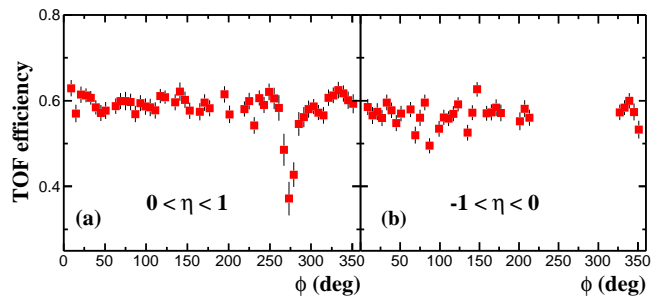}
\caption{(Color online) The TOF efficiency as a function of
azimuthal angle in the positive and negative $\eta$ region. The
hole in the negative $\eta$ region corresponds to the trays (28\%)
missing from the 2009 run installation. The 3 trays around 270
degree in the positive $\eta$ region have smaller matching
efficiencies but have similar performances in other aspects. }
\label{tofeff}
\end{center}
\end{figure}

Figure~\ref{eeff} shows the efficiency for single electrons in the
pseudo-rapidity range of $|\eta|\!<\!1$ in $p+p$ collisions at
$\sqrt{s} = 200$ GeV. Open circles represent the TPC tracking
efficiency alone. Including TOF matching and response decreases
the efficiency as shown by the triangles. With additional $dE/dx$
cuts, the final efficiency for single electrons for
$|\eta_{e}|\!<\!1$ is shown as squares. For this analysis, 72\%
out of a possible 120 total TOF trays were installed and the
efficiency dependence on azimuthal angle is shown in
Fig.~\ref{tofeff} for positive and negative $\eta$ regions. We
have accounted for the incomplete TOF acceptance in determining
efficiencies for the di-electron spectra.

\renewcommand{\floatpagefraction}{0.75}

\renewcommand{\floatpagefraction}{0.75}

\begin{figure*}[htbp]
\begin{center}
\includegraphics[keepaspectratio,width=0.8\textwidth]{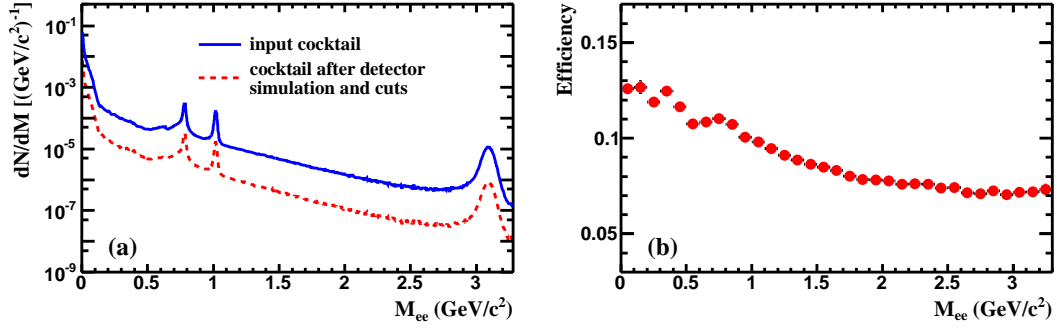}
\caption{(Color online) (a) The $M_{ee}$ distribution within
STAR's acceptance ($|y_{e^{+}e^{-}}|\!<\!1$, $|\eta_{e}|\!<\!1$,
and $p_T(e)\!>\!0.2$ GeV/$c$) from simulation. The solid line
represents the input cocktail. The dashed line represents the
cocktail from GEANT simulations taking into account the proper
efficiency factors described in the text. (b) The efficiency of
the di-electron continuum as a function of $M_{ee}$ within the
STAR acceptance in $p+p$ collisions at $\sqrt{s} = 200$ GeV. }
\label{dieeff}
\end{center}
\end{figure*}

The efficiency factor for the di-electron continuum within STAR's
acceptance is obtained in two steps. We obtain the input cocktail
{\it A} within STAR acceptance by the method described in
Sec.~\ref{cocktail}. The input cocktail includes the radiation
energy loss and momentum resolution determined from GEANT
simulations. The result is shown by the solid line in
Fig.~\ref{dieeff} (a). Then we obtain cocktail $B$ from GEANT
simulations with proper efficiency factors including the TPC
tracking, TOF acceptance and response, and $dE/dx$ cut for single
electrons as described above, and shown as the dashed line in
Figure~\ref{dieeff} (a). The ratio of these two is the efficiency
factor for the di-electrons shown in Fig.~\ref{dieeff} (b) for
$p+p$ collisions at $\sqrt{s} = 200$ GeV. The uncertainty on the
efficiency factor is about 10\% with a negligible $p_T$
dependence. The final di-electron continuum within the STAR
acceptance is obtained after this correction is applied and is
discussed in Sec.~\ref{correctedcontinuum}.

\subsection{Results}\label{correctedcontinuum}
\begin{figure}[htbp]
\begin{center}
\includegraphics[keepaspectratio,width=0.45\textwidth]{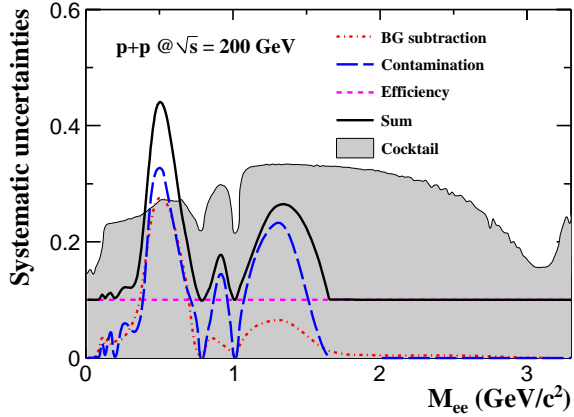}
\caption{(Color online) Systematic uncertainties as a function of
mass from different source contributions. Also shown are the
systematic uncertainties of the cocktail simulation.}
\label{sysdata}
\end{center}
\end{figure}

The systematic uncertainties on the di-electron continuum are
dominated by background subtraction (acceptance difference between
like-sign and unlike-sign electron pairs and normalization of
mixed-event distributions) and hadron contamination. The
acceptances of the like-sign and unlike-sign distributions are the
same within 5\% for $M_{ee}\!>\!0.1$ GeV/$c^{2}$ due to the
azimuthal symmetry and the solenoidal magnetic field. The small
acceptance differences due to track merging, sector boundaries,
and dead channels have been evaluated using the differences of the
unlike-sign and like-sign distributions from the mixed-event
technique. This difference is included in the systematic
uncertainty. In addition, for $M_{ee}\!>\!0.4$ GeV/$c^{2}$, the
normalization factor between the mixed-event and the like-sign
distributions contributes 0-7\% to the overall uncertainty. The
systematic uncertainty from efficiency factors is about 10\% with
no significant mass dependence. The uncertainties in hadron
contamination (hadrons from resonance decays mis-identified as
electrons) are 0-32\% and are mass dependent. Figure~\ref{sysdata}
shows the relative systematic uncertainties from different sources
for each mass bin. The normalization uncertainty in NSD $p+p$
collisions is 8\%~\cite{weihardprobes2010}. Additional
contribution from the normalization uncertainty in VPD triggered
minimum-bias events taking into account the trigger bias and
vertex finding efficiency is 8\% as determined from PYTHIA
simulations. The total normalization uncertainty for di-electron
mass spectra is 11\% in $p+p$ collisions. Table~\ref{tab:II} shows
the systematic uncertainties from different sources.
\renewcommand{\floatpagefraction}{0.75}

The uncertainties on the cocktail simulations include the decay
form factors and the measured cross section for each hadron. By
fitting the di-electron continuum, open charm~\cite{starelectron}
and non-photonic electron spectra~\cite{weihardprobes2010}
simultaneously, we find that the $c\bar{c}$ cross section in 200
GeV $p+p$ collisions is $0.92\pm0.10\pm0.26$ mb, consistent with
earlier RHIC measurements~\cite{starelectron,phenixelectron}. This
is used to generate the charm component in this paper. The
systematic uncertainties for the cocktail simulation as a function
of mass are shown in Fig.~\ref{sysdata}. Future precise
measurements of di-electron continuum in the intermediate mass
region can further constrain the charm production mechanism in
$p+p$ collisions.
\begin{table}\caption{Systematic uncertainties from different sources for di-electron continuum.\label{tab:II}}
\begin{tabular}{c|c} \hline\hline
\centering
 source & contribution factors\\
 \hline
background subtraction&0-27\%\\
contamination&0-32\%\\
efficiency &10\%\\
total normalization &11\%\\
cocktail simulation&14-33\%\\
\hline \hline
\end{tabular}

\end{table}

\begin{figure}[htbp]
\begin{center}
\includegraphics[keepaspectratio,width=0.45\textwidth]{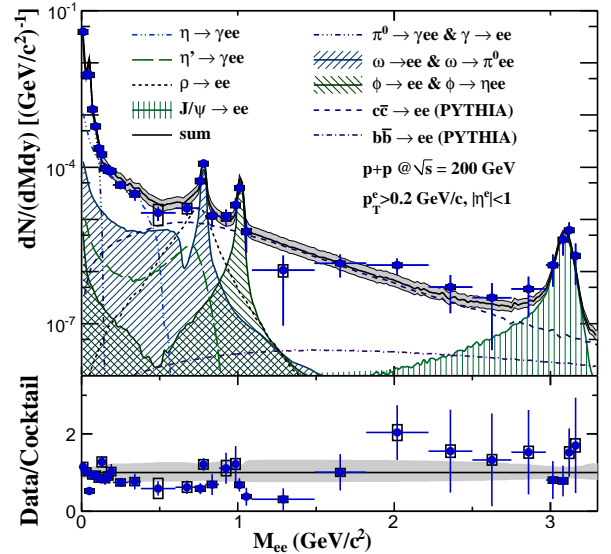}
\caption{(Color online) The comparison of the di-electron
continuum between data and simulation after efficiency correction
within the STAR acceptance in $\sqrt{s} = 200$ GeV NSD $p+p$
collisions. The di-electron continuum from simulations with
different source contributions is also shown. The statistical
errors on the data are shown as bars, while the systematic
uncertainties are shown as boxes. The 11\% normalization
uncertainty on the data is not shown. The band on top of the solid
curve illustrates the systematic uncertainties on the cocktail
simulation.} \label{continuum}
\end{center}
\end{figure}

After the efficiency correction, the di-electron continuum within
the STAR acceptance is shown in Fig.~\ref{continuum} for $\sqrt{s}
= 200$ GeV NSD $p+p$ collisions. The di-electron mass spectrum is
not corrected for momentum resolution and radiation energy loss
effects. The ratio of data to cocktail simulation is shown in the
lower panel of Fig.~\ref{continuum}. Within the uncertainties, the
cocktail simulation is consistent with the measured di-electron
continuum. The $\chi^2/NDF$ between data and cocktail simulation
are 21/26 for $M_{ee}\!>\!0.1$ GeV/$c^2$ and 8/7 for
$1.1\!<\!M_{ee}\!<\!3.0$ GeV/$c^2$. In the mass region
$0.2\!<\!M_{ee}\!<\!0.8$ GeV/$c^2$, the cocktail simulation is
systematically higher than the measured di-electron continuum.
However, they are also consistent within uncertainties. We find
that better agreement between the cocktail simulation and data can
be achieved by applying an additional scale factor (56\%) to the
$\eta$ Dalitz decay contribution. Further details on this decay
can be found in Sec.~\ref{raredecay}.

\renewcommand{\floatpagefraction}{0.75}

\section{Vector meson production}\label{vecresults} The yields of
the $\omega$, $\phi$ and $J/\psi$ long-lived vector mesons can be
extracted from the di-electron continuum. We use the mixed-event
technique to reconstruct the combinatorial background beneath the
respective peaks. The mixed-event distribution is normalized by a
constant to match the like-sign distribution in the mass region of
0.4-1.5 GeV/$c^{2}$, as discussed in Sec.~\ref{bg}. The background
is then subtracted to obtain the signal, which will still contain
some residual background of di-electron pairs from other sources
as described in Sec.~\ref{cocktail}.

\begin{figure*}[htbp]
\begin{center}
\includegraphics[keepaspectratio,width=0.8\textwidth]{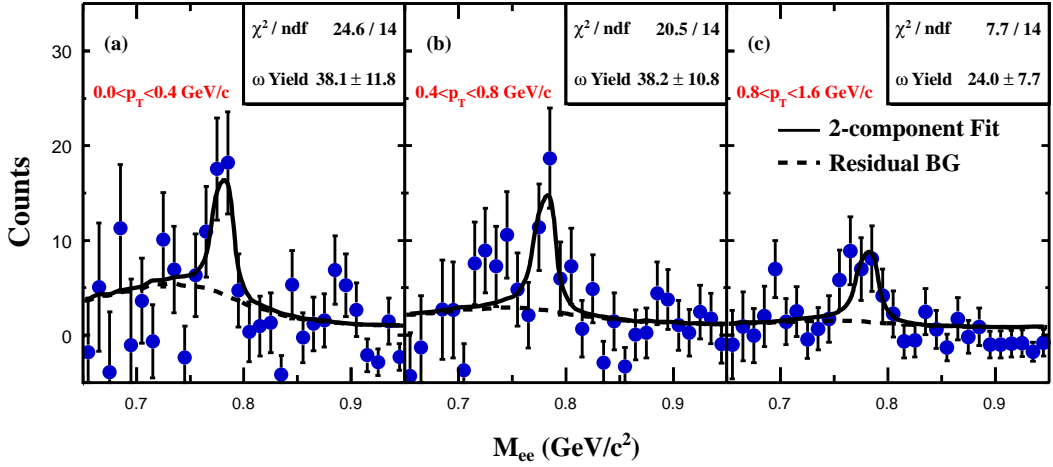}
\caption{(Color online) The $M_{ee}$ distribution in the range
$0.65\!<\!M_{ee}\!<\!0.95$ GeV/$c^{2}$ for three different $p_T$
bins after the mixed-event background subtraction in $p+p$
collisions at $\sqrt{s} = 200$ GeV. The curves represent fits in
the range $0.7\!<\!M_{ee}\!<\!0.85$ GeV/$c^{2}$ as described in
the text. Errors on data points are statistical.} \label{omegafit}
\end{center}
\end{figure*}

A two-component fit is used to extract the raw signal $\omega
\rightarrow e^{+}e^{-}$ from the residual background in the
invariant mass range of $0.7\!<\!M_{ee}\!<\!0.85$ GeV/$c^{2}$. The
first component represents the line shape (the $\omega \rightarrow
e^{+}e^{-}$ signal shape), the second the residual background. The
line shape of the $\omega \rightarrow e^{+}e^{-}$ invariant mass
distribution, and the shape and magnitude of the background are
determined from the simulations described in Sec.~\ref{cocktail}.
The systematic uncertainties of the $\omega\rightarrow e^{+}e^{-}$
raw yields are derived by changing the magnitude of the background
allowed by the uncertainties of the cocktail simulation and the
analysis cuts. The total contribution to the raw yield is about
20\%. Figure~\ref{omegafit} shows the fit to the $M_{ee}$
distribution for $0.7\!<\!M_{ee}\!<\!0.85$ GeV/$c^{2}$ in three
different $p_T$ bins for $p+p$ collisions at $\sqrt{s} = 200$ GeV.
\renewcommand{\floatpagefraction}{0.75}

\renewcommand{\floatpagefraction}{0.75}
\begin{figure}[htbp]
\begin{center}
\includegraphics[keepaspectratio,width=0.4\textwidth]{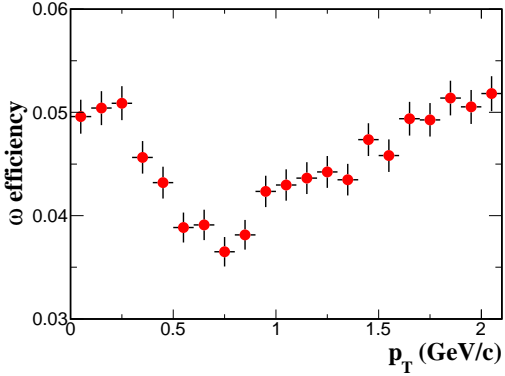}
\caption{(Color online) The efficiency including STAR acceptance
for $\omega \rightarrow e^{+}e^{-}$ for $|y|\!<\!1$ in $p+p$
collisions at $\sqrt{s} = 200$ GeV.} \label{omegaeff}
\end{center}
\end{figure}

\begin{figure}[htbp]
\begin{center}
\includegraphics[keepaspectratio,width=0.45\textwidth]{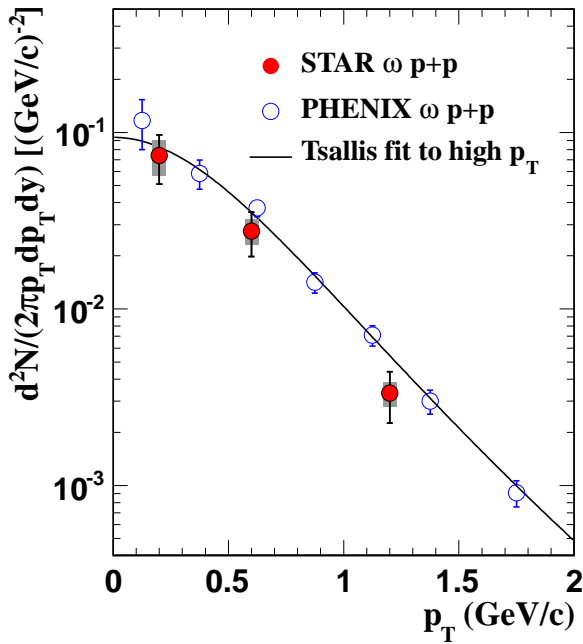}
\caption{(Color online) The $\omega \rightarrow e^{+}e^{-}$
invariant yield, divided by its B.R., in NSD $p+p$ collisions at
$\sqrt{s} = 200$ GeV. The open circles represent PHENIX published
results~\cite{phenixomegappII}. The bars are statistical errors
and boxes are systematic uncertainties. A normalization
uncertainty of 11\% is not shown. The yields at the center of the
$p_T$ bin are corrected for finite bin width. The line represents
the yields of $\omega$ from Tsallis function fit to high $p_T$
$\omega$ yields measured from its hadronic decay and described in
Sec.~\ref{cocktail}.} \label{omegayield}
\end{center}
\end{figure}

\renewcommand{\floatpagefraction}{0.75}
In order to present the final differential invariant cross section
as a function of $p_T$, the raw vector meson yield ($\omega
\rightarrow e^{+}e^{-}$) is corrected for acceptance and all
detector effects which reduce the measured raw yield relative to
the actual yield. The total efficiency correction for $\omega
\rightarrow e^{+}e^{-}$ for $|y|\!<\!1$ is shown in
Fig.~\ref{omegaeff}. We use the simulations described in
Sec.~\ref{cocktail} to determine this correction, which accounts
for limits in TPC acceptance and inefficiencies in TPC tracking,
limits in the TOF acceptance/response, and the rejection of signal
due to the $dE/dx$ cut. The invariant yield is defined as follows:
\begin{equation}
\frac{d^2N}{2{\pi}p_Tdp_Tdy}=\frac{N_{raw}\times
NORM}{2{\pi}p_Tdp_Tdy \times N_{event} \times \epsilon \times
B.R.},
\end{equation}
where $N_{raw}$ represents the raw signal counts, $N_{event}$ is
the total event number, $\epsilon$ is the total efficiency and
acceptance correction factor, $B.R.$ is the branching ratio for
$\omega \rightarrow e^{+}e^{-}$, and $NORM$ is the normalization
factor ($64\pm5$\% for VPD triggered minimum-bias events taking
into account the trigger bias and vertex finding efficiency,
determined by, and corrected for, using PYTHIA simulations). The
$\omega$ invariant yield in NSD $p+p$ collisions at $\sqrt{s}
=200$ GeV is presented in Fig.~\ref{omegayield}. The systematic
uncertainties are dominated by uncertainties in the two-component
fit, and uncertainties in total efficiency which are also
described in Sec.~\ref{cocktail}. Table~\ref{omegasys} lists the
detailed systematic uncertainties for $\omega$ yields from
different sources. Our $\omega$ yields from di-electron decays are
consistent with previous results~\cite{phenixomegappII} and with a
prediction from a Tsallis fit, which fits spectra of other
particles and high $p_T$ $\omega$
yields~\cite{Tsallis,phenixomegapp}. We obtain a mid-rapidity
$dN/dy$ of 0.10 $\pm$ 0.02 (stat.) $\pm$ 0.02 (sys.) for the
$\omega$.

\begin{table}\caption{Systematic uncertainties from different sources for $\omega$ yields.\label{omegasys}}
\begin{tabular}{c|c} \hline\hline
\centering
 source & contribution factors\\
 \hline
two component fit&9-16\%\\
$n\sigma_{e}$ cut&3-7\%\\
efficiency &10\%\\
total normalization &11\%\\
\hline \hline
\end{tabular}

\end{table}

\begin{figure*}[htbp]
\begin{center}
\includegraphics[keepaspectratio,width=0.90\textwidth]{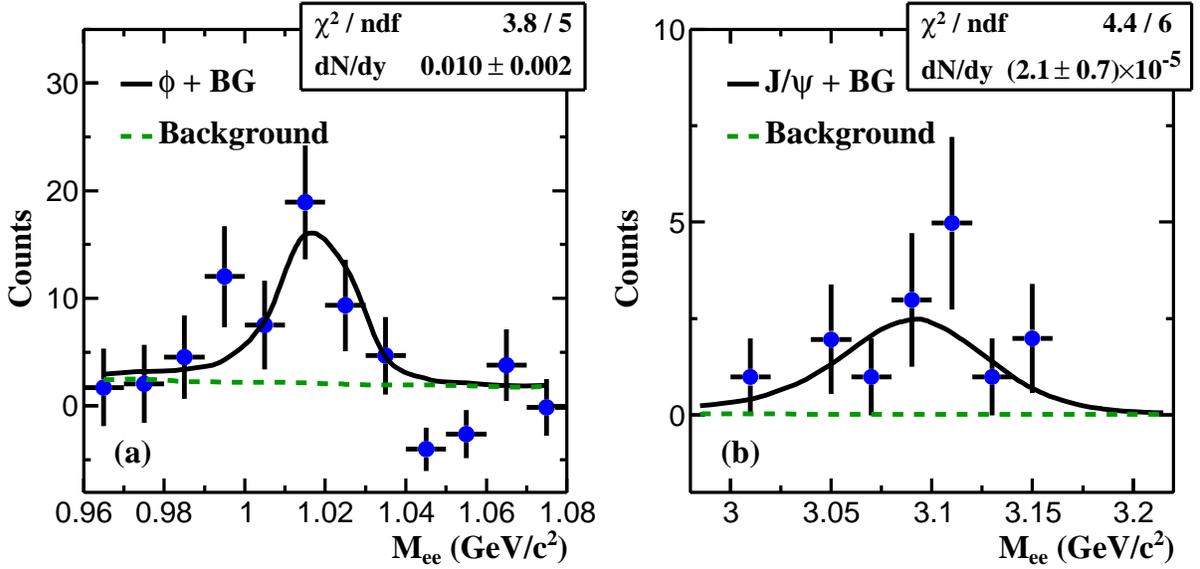}
\caption{(Color online) Panel (a): The $M_{ee}$ distribution after
the mixed-event background subtraction for
$0.96\!<\!M_{ee}\!<\!1.08$ GeV/$c^{2}$ in $p+p$ collisions at
$\sqrt{s} = 200$ GeV. The curve represents the fit in the range of
$0.98\!<\!M_{ee}\!<\!1.04$ GeV/$c^{2}$. Errors on data points are
statistical. Panel (b): The $M_{ee}$ distribution after the
mixed-event background subtraction for $2.98\!<\!M_{ee}\!<\!3.22$
GeV/$c^{2}$ in $p+p$ collisions at $\sqrt{s} = 200$ GeV. The curve
represents the fit in the range of $3\!<\!M_{ee}\!<\!3.16$
GeV/$c^{2}$. There is no count in the unlike-sign and like-sign
electron-pair distributions for $3.02\!<\!M_{ee}\!<\!3.04$
GeV/$c^{2}$ due to statistical fluctuations. Errors on data points
are statistical.} \label{phifit}
\end{center}
\end{figure*}

In addition to $\omega \rightarrow e^{+}e^{-}$ yields, we also
obtain the mid-rapidity yields, $dN/dy$, for the $\phi$ and
$J/\psi$ particles in NSD $p+p$ collisions at $\sqrt{s} =200$ GeV.
Due to limited statistics for each particle, the invariant mass
signal can only be extracted over all $p_T$, rather than
individual bins. As before, we use a two-component fit in the
region of $M_{ee}$ for $0.98\!<\!M_{ee}\!<\!1.04$ GeV/$c^{2}$ for
the $\phi\rightarrow e^{+}e^{-}$, and $3\!<\!M_{ee}\!<\!3.16$
GeV/$c^{2}$ for $J/\psi\rightarrow e^{+}e^{-}$. The line shapes of
$\phi \rightarrow e^{+}e^{-}$ and $J/\psi\rightarrow e^{+}e^{-}$
are from simulations as discussed in Sec.~\ref{cocktail}. The
residual background shape and magnitude are obtained from the
cocktail simulation, shown in Sec.~\ref{cocktail}. The systematic
uncertainties of the $\phi$ and $J/\psi$ $dN/dy$ due to the
residual background are evaluated by changing the magnitude of the
background allowed by the uncertainties of the cocktail
simulation. The total efficiency correction for each particle is
evaluated in the same way as for the $\omega \rightarrow
e^{+}e^{-}$ analysis. Since the correction depends on $p_T$, we
calculate a weighted average using the predicted spectra from the
previously mentioned Tsallis fit as $p_{T}$ weights. The total
efficiency and acceptance correction factors for $\phi$ and
$J/\psi$ are 4.4\% and 3.4\%, respectively. The systematic
uncertainty on the total efficiency correction is estimated to be
10\%. The normalization uncertainty is 11\% in $p+p$ collisions.
Table~\ref{phisys} lists the detailed systematic uncertainties for
the $\phi$ and $J/\psi$ $dN/dy$ from different sources.
Figure~\ref{phifit} shows the fits to the $M_{ee}$ distributions
used to obtain the mid-rapidity yields $dN/dy$ for $\phi$ and
$J/\psi$ in NSD $p+p$ collisions at $\sqrt{s}= 200$ GeV. The final
yields in those fits are subject to the total efficiency
correction. The $dN/dy$ for $\phi$ is 0.010 $\pm$ 0.002 (stat.)
$\pm$ 0.002 (syst.), consistent with the measurements from
$\phi\rightarrow K^{+}K^{-}$~\cite{starphipp,phenixphipp}. The
$dN/dy$ for $J/\psi$ is (2.1 $\pm$ 0.7 (stat.) $\pm$ 0.7 (syst.))
$\times$ $10^{-5}$, consistent with previous
measurements~\cite{phenixjpsipp,starjpsipp}.

\begin{table}\caption{Systematic uncertainties from different sources for $\phi$ and $J/\psi$ $dN/dy$.\label{phisys}}
\begin{tabular}{c|c|c} \hline\hline
\centering
 source & contribution factors for $\phi$ & for $J/\psi$\\
 \hline
two component fit&10\% & 7\%\\
$n\sigma_{e}$ cut&8\% &27\%\\
efficiency &10\% &10\%\\
total normalization &11\% &11\%\\
\hline \hline
\end{tabular}
\end{table}


\renewcommand{\floatpagefraction}{0.75}


\section{The rare decay: $\eta\rightarrow e^{+}e^{-}$}\label{raredecay}
\begin{figure}[htbp]
\begin{center}
\includegraphics[keepaspectratio,width=0.45\textwidth]{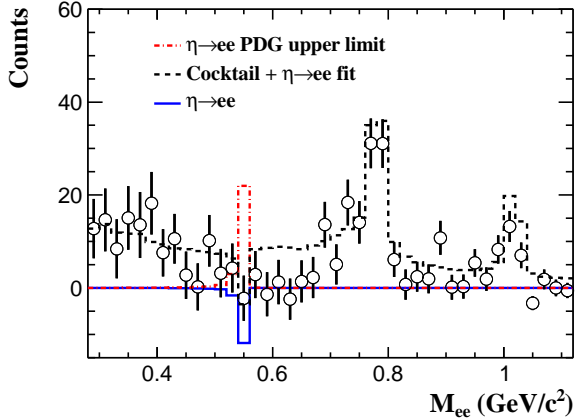}
\caption{(Color online) $M_{ee}$ distribution for di-electron
production in $p+p$ collisions at $\sqrt{s}=200$ GeV. The dashed
curve is the cocktail with the $\eta \rightarrow e^+e^-$ decay
channel included in the fit. The $\eta \rightarrow e^+e^-$
contribution is shown as the solid curve. The dot-dashed peak is
the $\eta \rightarrow e^+e^-$ contribution with the upper limit of
its branching ratio from the PDG~\cite{CLEOeta}, which is
$2.7\times10^{-5}$.} \label{etalim}
\end{center}
\end{figure}

As discussed in previous sections and shown in
Fig.~\ref{continuum}, the cocktail can describe the data
reasonably well around the $\eta$ mass without the $\eta
\rightarrow e^+e^-$ decay channel. We zoom into the low-mass range
and show the data and cocktail comparison in Fig.~\ref{etalim}.
The dot-dashed peak is the $\eta \rightarrow e^+e^-$ contribution
with the upper limit of its B.R. from the PDG~\cite{pdg}, which is
$2.7\times10^{-5}$~\cite{CLEOeta}. The dashed curve is a
two-component fit with the $\eta \rightarrow e^+e^-$ decay channel
included. The fit function is
\begin{equation} A\times N_{\eta} + B\times Cocktail,
\end{equation} in which $N_{\eta}$ is
the expected $\eta$ contribution with the line shape of $\eta
\rightarrow e^+e^-$ after detector simulation, $Cocktail$ is the
expected cocktail contribution described in Sec.~\ref{cocktail}
without $\eta \rightarrow e^+e^-$, and $A$ and $B$ are fit
parameters. The $A$ and $B$ represent the B.R. for the $\eta
\rightarrow e^+e^-$ and a scale factor for the cocktail
contribution, respectively. The solid curve shown in
Fig.~\ref{etalim} represents the $\eta \rightarrow e^+e^-$
contribution from the fit. It gives the B.R. of the $\eta
\rightarrow e^+e^-$ to be ($-$9.6 $\pm$ 5.9 (stat.) $\pm$ 5.3
(syst.)) $\times$ $10^{-6}$. The negative value of B.R. of the
$\eta \rightarrow e^+e^-$ is due to the statistical fluctuation at
$M_{ee}=0.55$ GeV/$c^{2}$. The systematic uncertainties are
dominated by background subtraction (34\%, the difference between
mixed event background subtraction and like-sign subtraction),
electron purity (31\%, determined by varying the $n\sigma_e$
cuts), and track quality cut (27\%, determined by changing the cut
of distance of closest approach between the track and the
collision vertex).

In addition, although the Dalitz decay yield of
$\eta\rightarrow\gamma e^{+}e^{-}$ is consistent with the cocktail
expectation from the Tsallis fit to $\eta$ measurements for
$p_T\!>\!2$ GeV/$c$ described in Sec.~\ref{cocktail}, the nominal
value from the fit in Fig.~\ref{etalim} is about 56\% of the input
cocktail ($B$=56\%). This additional factor leads to a better
agreement between data and cocktail simulation compared to that
shown in Fig.~\ref{continuum}. This is equivalent to a smaller
value of $N_{\eta}$ and has to be taken into account when the
upper limit at the 90\% Confidence Level (CL) for the B.R. of
$\eta \rightarrow e^+e^-$ is estimated.

With different background subtraction, electronic purity, and
track quality cuts, we repeat the fit process described above to
obtain the parameter $A$, the B.R. of the $\eta \rightarrow
e^+e^-$. The difference of the $A$ values is attributed to the
systematic uncertainties. In this fit procedure, we find that cut
conditions do not contribute to the point-to-point variation
around $\eta$ mass range and the statistical error of the
parameter $A$ remains unchanged. Therefore, the significance of an
observable signal above the background only depends on the
statistical fluctuation. To estimate the upper limit for the B.R.
of $\eta \rightarrow e^+e^-$ at the 90\% CL due to a possible
statistical fluctuation, we utilize the statistical error
$5.9\times10^{-6}$ for the B.R. of the $\eta \rightarrow e^+e^-$
from the fit and set the lower bound of the B.R. of $\eta
\rightarrow e^+e^-$ to be zero instead of the negative value from
the fit. The upper limit is found to be 5.9 $\times$ $10^{-6}$
$\times$ 1.65/0.56 = 1.7 $\times$ $10^{-5}$, in which 1.65 is the
upper endpoint of a confidence interval (the lower endpoint is
zero) with a 90\% CL for a standard normal distribution.


These results provide a promising first glimpse of a program on
searching for rare decays of hadrons produced in relativistic
heavy-ion collisions at STAR. With large hadron yields, high
efficiency for electrons at low momentum and high mass resolution,
STAR provides a unique tool for such a program in the years to
come.

\section{Summary}\label{summary} The di-electron continuum is
measured in $\sqrt{s}=200$ GeV non-singly diffractive $p+p$
collisions within the STAR acceptance. The cocktail simulations
are consistent with the data and provide a reference for the
future heavy-ion studies. The $\omega$ invariant yields are
consistent with the previous publications. The $dN/dy$ for $\phi$
and $J/\psi$ are 0.010 $\pm$ 0.002 (stat.) $\pm$ 0.002 (syst.) and
(2.1 $\pm$ 0.7 (stat.) $\pm$ 0.7 (syst.)) $\times$ $10^{-5}$,
respectively. These results are consistent with the previous
measurements from $\phi\rightarrow K^{+}K^{-}$ and
$J/\psi\rightarrow e^{+}e^{-}$. Our measurement lowers the current
world limit of the branching ratio of the $\eta \rightarrow
e^{+}e^{-}$ from $2.7\times10^{-5}$ to $1.7\times10^{-5}$.

We thank the RHIC Operations Group and RCF at BNL, the NERSC
Center at LBNL and the Open Science Grid consortium for providing
resources and support. This work was supported in part by the
Offices of NP and HEP within the U.S. DOE Office of Science, the
U.S. NSF, the Sloan Foundation, the DFG cluster of excellence
`Origin and Structure of the Universe' of Germany, CNRS/IN2P3,
FAPESP CNPq of Brazil, Ministry of Ed. and Sci. of the Russian
Federation, NNSFC, CAS, MoST, and MoE of China, GA and MSMT of the
Czech Republic, FOM and NWO of the Netherlands, DAE, DST, and CSIR
of India, Polish Ministry of Sci. and Higher Ed., Korea Research
Foundation, Ministry of Sci., Ed. and Sports of the Rep. of
Croatia, and RosAtom of Russia.








\end{document}